# Transitional random matrix theory nearest-neighbor spacing distributions

**Fredy R Zypman**

Department of Physics, Yeshiva University, New York, NY, USA

**Abstract**

This paper presents a study of the properties of a matrix model that was introduced to describe transitions between all Wigner surmises of Random Matrix theory. New results include closed-form exact analytical expressions for the transitional probability density functions, as well as suitable analytical approximations for cases not amenable to explicit representation.

**Introduction**

In this paper we find analytical expressions for probability density functions (pdf) that are hybrid between Wigner surmises of Random Matrix theory (RMT) of the Gaussian unitary ensemble, the Gaussian orthogonal ensemble and the Gaussian symplectic ensemble, as well as the Ginibre pdf. This is done by exploiting the connection between the level spacing statistics of a matrix described below and the Gaussian distribution of points in hyperspace.

Much of the work to find pdfs in RMT is done, by necessity, numerically. However, simple analytical results are habitually more instructive. They can expose the pdfs under study in a very different light from that provided by specific numerical examples. That is the justification of this work.

The statistical properties of the eigenvalues of random matrices have shown to be valuable in the modeling of a broad variety of natural systems. After its seminal inception [1], the subject flourished after Wigner introduced statistical concepts to study nuclear energy levels, and eventually ensembles of random matrices [2,3,4]. Dyson classified the random matrix ensembles according to their invariance properties under time reversal [5,6,7,8,9]. For Hermitian random matrices the invariant Gaussian ensembles are the Gaussian unitary ensemble (GUE), the Gaussian orthogonal ensemble (GOE) and the Gaussian symplectic ensemble (GSE).

RMT as been used to analyze a vast array of physical phenomena involving discrete levels. In those cases, it is customary to obtain the nearest-neighbor spacing distribution (NNSD) of levels, and compare it to results from RMT [10]. Typically, this comparison is done against the Poisson distribution, or the analytical Wigner surmises of RMT probability density functions,

$$P_W(s) = C_1 s^\beta e^{-C_2 s^2} \qquad (1)$$

where $C_1$ and $C_2$ are found from normalization and $\langle s \rangle = 1$. The GOE surmise corresponds to $\beta = 1$, GUE to $\beta = 2$, and GSE to $\beta = 4$ [11]. The value $\beta = 3$ corresponds to the known Ginibre ensemble, which has been used to study the statistical properties of the roots of random polynomials [12].

It is common to find that the NNSD of levels is not described by the above pure distributions, but rather by hybrids between them [13,14,15,16,17]. Studies of these hybrid distributions have been made in connection with 2x2 and 4x4 random matrices [18,19] that, in appropriate limits, are shown to have NNSD of levels given by (1) with $\beta = 1,2,3,4$. In addition, it was shown in references 18 and 19 that transitional NNSD can be obtained by appropriately choosing the parameters of the model matrices.

In this work we consider the 4x4 random matrices of reference 18, and provide closed-form analytical expressions for some of the possible transitional distributions.

**Theoretical Preliminaries**

To implement the ideas introduced above, we consider a specific matrix whose energy levels spacing statistics are pure Wigner surmises of RMT (plus Ginibre) and all the intermediate possibilities.

Specifically, we begin with the matrix [18],

$$H = \begin{pmatrix} a & 0 & c & 0 \\ 0 & a & 0 & c \\ c & 0 & b & 0 \\ 0 & c & 0 & b \end{pmatrix} + i\alpha_1 \begin{pmatrix} 0 & 0 & d & 0 \\ 0 & 0 & 0 & -d \\ -d & 0 & 0 & 0 \\ 0 & d & 0 & 0 \end{pmatrix} + \alpha_2 \begin{pmatrix} 0 & 0 & 0 & e \\ 0 & 0 & -e & 0 \\ 0 & -e & 0 & 0 \\ e & 0 & 0 & 0 \end{pmatrix} + i\alpha_3 \begin{pmatrix} 0 & 0 & 0 & f \\ 0 & 0 & f & 0 \\ 0 & -f & 0 & 0 \\ -f & 0 & 0 & 0 \end{pmatrix} \quad (2)$$

where a, b, c, d, e and f are Gaussian distributed random variables of mean zero and variance 2,2,1,1,1 and 1 respectively. Each $\alpha$ is a real parameter in the range $0 \leq \alpha \leq 1$.

One might inquire, at this point, about the relevance of matrix (2) to RMT. This question arises because in RMT one is interested in results for $\hat{N} \times \hat{N}$ matrices with $\hat{N}$ large, while H is a $4 \times 4$ matrix. There are two answers that justify a detailed study of (2). First, it is useful, when programming pdfs for large matrices in RMT, to have known limiting cases against which to test the code. Second, because it is common to compare level spacing statistics not to the exact RMT results, but to Wigner surmises, which are known to be excellent analytical approximations to the level statistics for Gaussian ensembles of large random matrices [20,21,22]. Thus similarly, finding analytical pdfs for the transitional level statistics for matrix (2) could be useful when studying level statistics for large random matrices that might fall into the transitional regime.

For each $(\alpha_1 \quad \alpha_2 \quad \alpha_3)$, the eigenvalues of H are double degenerate. The near neighbor spacing is

$$s = 2\sqrt{g^2 + c^2 + \alpha_1^2 d^2 + \alpha_2^2 e^2 + \alpha_3^2 f^2} \qquad (3)$$

where $2g = a - b$ and consequently has zero mean and variance 1.

A slight generalization of formulas given in reference (19) shows that the probability density function of $\frac{s}{2}$ is obtained by randomly sampling an N-dimensional (the limiting cases are N=5 if all $\alpha$ are nonzero, and N=2 if all are zero) space using the following probability density function,

$$\Omega(g,c,d,e,f;\alpha_1,\alpha_2,\alpha_3) = \frac{1}{2\pi}\left[\prod_{i=1}^{M+2}\left(\frac{1}{\sqrt{2\pi}\tilde{\alpha}_i}\right)\right]\exp\left[-\frac{1}{2}\sum_{i=1}^{M+2}\left(\frac{x_i}{\tilde{\alpha}_i}\right)^2\right] \quad (4)$$

where $x_1 = g$ etcetera, $\tilde{\alpha}_1 = \tilde{\alpha}_2 = 1$, $\tilde{\alpha}_3 = \alpha_1$, $\tilde{\alpha}_4 = \alpha_2$, $\tilde{\alpha}_5 = \alpha_3$ (unless some of the $\alpha_i$ vanish as explained below) and M is the number of $\alpha$ that are nonzero (if $\alpha_j = 0$, the corresponding variable in equation (3) does not contribute, and the dimensionality N of the space is reduced in one unit).

For example, if no $\alpha$ vanishes then M=3 and the vector $\vec{x}$ lives in 5-dimensional space. In that case, the coordinates can be parametrized in hyperspherical coordinates [23],

$$\begin{cases} f = x_5 = \frac{s}{2}\cos\xi \\ e = x_4 = \frac{s}{2}\sin\xi\cos\psi \\ d = x_3 = \frac{s}{2}\sin\xi\sin\psi\cos\theta \qquad 0 \leq \varphi \leq 2\pi \quad 0 \leq \theta,\psi,\xi \leq \pi \\ c = x_2 = \frac{s}{2}\sin\xi\sin\psi\sin\theta\cos\varphi \\ g = x_1 = \frac{s}{2}\sin\xi\sin\psi\sin\theta\sin\varphi \end{cases} \quad (5)$$

then, the desired probability density function F, corresponds to all points $\vec{x}$ in a thin shell around $\frac{s}{2}$, namely $\frac{s}{2} \leq |\vec{x}| \leq \frac{s}{2} + d(\frac{s}{2})$ (where $d$ here represents a differential, not to be confused with the parameter introduced in (2)),

$$F(s,\alpha_1,\alpha_2,\alpha_3)ds = \frac{d(s/2)}{(2\pi)^{5/2}\alpha_1\alpha_2\alpha_3} \int_0^{2\pi}\int_0^{\pi}\int_0^{\pi}\int_0^{\pi} \left[\left(\frac{s}{2}\right)^4 \sin^3\xi \sin^2\psi \sin\theta\, d\xi\, d\psi\, d\theta\, d\varphi\right]$$
$$\times \exp\left[-\left(\frac{s}{2}\right)^2 \left(\sin^2\xi \sin^2\psi \sin^2\theta + \frac{\sin^2\xi \sin^2\psi \cos^2\theta}{\alpha_1^2} + \frac{\sin^2\xi \cos^2\psi}{\alpha_2^2} + \frac{\cos^2\xi}{\alpha_3^2}\right)\right]$$
(6).

The expression inside the first square bracket corresponds to the volume of the thin shell of width $ds/2$. Similar expressions are obtained when two $\alpha$ are zero (integral in 3D space), and when only one $\alpha$ vanishes (integral in 4D space). In all cases, the integral in $\varphi$ equals $2\pi$ and thus the problem reduces to performing a triple, double, or single integral.

From (6) and the similar expressions in less dimensions, one readily finds what happens in the limiting cases of $\alpha = 0$ or $1$. For $\vec{\alpha} = (000)$, $F(s)$ gives rise to $\beta = 1$ (GOE) in equation (1), $\vec{\alpha} = (100)$ to $\beta = 2$ (GUE), $\vec{\alpha} = (110)$ to $\beta = 3$ (Ginibre), and $\vec{\alpha} = (111)$ to $\beta = 4$ (GSE) (the last case is by direct inspection of equation (6)).

**Transitional pdf GUE To Ginibre**

For intermediate values of $\alpha$ ($0 < \alpha < 1$) the integrals are generally cumbersome and not necessarily amenable of closed-form solutions.

However, we will show in this section how to find an explicit analytical expression for the transisitional pdf GUE to Ginibre.

As explained in the section Theoretical Preliminaries, the transition from GUE to Ginibre is given by the probability density function (this has already been derived in [18]),

$$F(s,\alpha) = \frac{s^3}{16\pi\alpha} \int_0^\pi e^{-\frac{s^2}{8}\left(\sin^2\psi + \frac{\cos^2\psi}{\alpha^2}\right)} \sin^2\psi \, d\psi \qquad (7),$$

which can be rewritten as

$$F(s,\alpha) = \frac{s^3}{16\pi\alpha} e^{-\frac{s^2}{8\alpha^2}} \int_0^\pi e^{\frac{s^2}{8}\left(\frac{1-\alpha^2}{\alpha^2}\right)\sin^2\psi} \sin^2\psi \, d\psi \quad (8).$$

Thus the problem reduces to finding the integral

$$\Im(\mu) = \int_0^\pi e^{\mu \sin^2\psi} \sin^2\psi \, d\psi \qquad (9),$$

where we have defined the parameter $\mu = \frac{s^2}{8}\left(\frac{1-\alpha^2}{\alpha^2}\right)$.

It is immediate that,

$$\Im(\mu) = 2\frac{d}{d\mu}\int_0^{\pi/2} e^{\mu \sin^2\psi} d\psi = 2\frac{d}{d\mu}\int_0^{\pi/2} e^{\frac{\mu}{2}(1-\cos 2\psi)} d\psi = \frac{d}{d\mu}\left[e^{\frac{\mu}{2}}\int_0^\pi e^{-\frac{\mu}{2}\cos\tau} d\tau\right] \qquad (10)$$

The integral in the square bracket is [24] $\pi I_0(\mu/2)$, then

$$\Im(\mu) = \frac{\pi}{2} e^{\mu/2}\left[I_0(\mu/2) + I_1(\mu/2)\right] \qquad (11),$$

where $I_n$ are modified Bessel functions [24], and we have used the properties of their derivatives.

Thus, the transitional (GUE to Ginibre) pdf, equation (8) is

$$F(s,\alpha) = \frac{s^3}{32\alpha} e^{-\frac{s^2}{16}\left(\frac{1+\alpha^2}{\alpha^2}\right)}\left[I_0\left(\frac{s^2}{16}\left(\frac{1-\alpha^2}{\alpha^2}\right)\right) + I_1\left(\frac{s^2}{16}\left(\frac{1-\alpha^2}{\alpha^2}\right)\right)\right] \qquad (12)$$

One is typically interested in the pdf for $\Re \equiv \frac{s}{\langle s \rangle}$, where $\langle s \rangle$ is the mean of the pdf in equation (12). To that end we need to compute,

$$\langle s \rangle = \int_0^\infty s F(s,\alpha) ds \qquad (13).$$

Instead of using equation (12) directly, we go back to equation (8) and write the mean explicitly as

$$\langle s \rangle = \int_0^\infty s ds \frac{s^3}{16\pi\alpha} e^{-\frac{s^2}{8\alpha^2}} \int_0^\pi e^{\frac{s^2}{8}\left(\frac{1-\alpha^2}{\alpha^2}\right)\sin^2\psi} \sin^2\psi \, d\psi \qquad (14)$$

The integral in $s$ is of the form $\int_0^\infty s^4 e^{-\Gamma s^2} ds$, which can be written immediately as $\frac{3\sqrt{\pi}}{8\Gamma^{5/2}}$. Thus, (14) reduces to

$$\langle s \rangle = \frac{3\sqrt{8}\alpha^4}{2\sqrt{\pi}} \int_0^\pi \frac{\sin^2\psi}{[1-(1-\alpha^2)\sin^2\psi]^{5/2}} d\psi = \frac{3\sqrt{8}\alpha^4}{\sqrt{\pi}} \int_0^{\pi/2} \frac{\sin^2\psi}{[1-\eta\sin^2\psi]^{5/2}} d\psi \qquad (15),$$

where we called $\eta = 1-\alpha^2$ in the last integral. Then,

$$\langle s \rangle = \frac{3\sqrt{8}\alpha^4}{\sqrt{\pi}} \left\{ \frac{2}{3} \frac{d}{d\eta} \int_0^{\pi/2} \frac{d\psi}{[1-\eta\sin^2\psi]^{3/2}} \right\} \qquad (16).$$

The integral in equation (16) reduces to the complete elliptic one of the first kind, $E(\eta)$ [25] so that,

$$\langle s \rangle = \frac{4\sqrt{2}\alpha^4}{\sqrt{\pi}} \frac{d}{d\eta}\left[\frac{E(\eta)}{1-\eta}\right] \qquad (17),$$

with the notation of AMS-55 [24].

Using reference [26] $2\eta \frac{dE}{d\eta} = E(\eta) - K(\eta)$, where $K(\eta)$ is the complete elliptic integral of the second kind, equation (17) becomes

$$\langle s \rangle = \frac{4\sqrt{2}\alpha^4}{\sqrt{\pi}} \frac{(1+\eta)E(\eta) - (1-\eta)K(\eta)}{2\eta(1-\eta)^2} = \frac{2\sqrt{2}}{\sqrt{\pi}} \frac{(2-\alpha^2)E(1-\alpha^2) - \alpha^2 K(1-\alpha^2)}{1-\alpha^2} \quad (18)$$

Finally, using equation (12),

$$F(\Re,\alpha) = \langle s \rangle \frac{(\langle s \rangle \Re)^3}{32\alpha} e^{-\frac{(\langle s \rangle \Re)^2}{16}\left(\frac{1+\alpha^2}{\alpha^2}\right)} \left[ I_0\left(\frac{(\langle s \rangle \Re)^2}{16}\left(\frac{1-\alpha^2}{\alpha^2}\right)\right) + I_1\left(\frac{(\langle s \rangle \Re)^2}{16}\left(\frac{1-\alpha^2}{\alpha^2}\right)\right) \right] \quad (19)$$

where $<s>$ is explicitly given in equation (18).

Figure 1 shows plots of equation (19) for $0 \leq \alpha \leq 1$. As expected, the slope at the origin decreases as $\alpha$ increases (corresponding to the transition from $\sim \Re^2$ to $\sim \Re^3$). There is a substantial change in the value at the peak (about 20%), although the position of the peak changes only slightly.

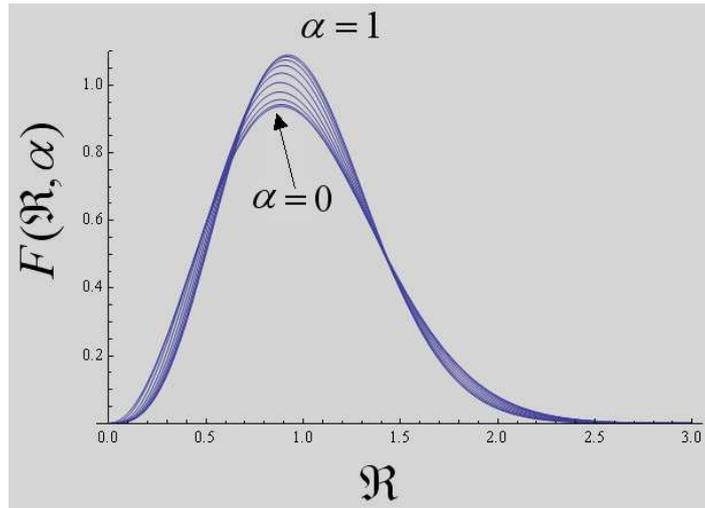

Figure 1. The transitional GUE to Ginibre pdfs $F(\Re,\alpha)$ have norm and mean one. Here shown are the pdfs for values of $\alpha$ ranging from 0 to 1.

**Transitional pdf Ginibre To GSE**

As mentioned in the previous section, intermediate values of $\alpha$ ($0 < \alpha < 1$) not necessarily render integrals that can be expressed in closed-form. However, similarly to what we did in that section, we will develop an explicit analytical expression for the transisitional pdf Ginibre to GSE.

Following the Theoretical Preliminaries, the transition from Ginibre to GSE is given by the probability density function (this has already been derived in [18]),

$$F(s,\alpha) = \frac{s^4}{64\sqrt{2\pi}\alpha} \int_0^\pi e^{-\frac{s^2}{8}\left(\sin^2\psi + \frac{\cos^2\psi}{\alpha^2}\right)} \sin^3\psi \, d\psi \qquad (20).$$

This integral can be solved by the change of variables $x = \cos\psi$,

$$\begin{aligned}
F(s,\alpha) &= \frac{s^4}{32\sqrt{2\pi}\alpha} \int_0^1 e^{-\frac{s^2}{8}\left(1-x^2+\frac{x^2}{\alpha^2}\right)} (1-x^2) dx \\
&= \frac{s^4 e^{-\frac{s^2}{8}}}{64\sqrt{2\pi}\alpha} \left\{ \frac{e^{-\mu}}{\mu} + \frac{\sqrt{\pi}(2\mu-1) Erf\left(\sqrt{\mu}\right)}{2\mu^{3/2}} \right\}
\end{aligned} \qquad (21),$$

where, $Erf$ is the error function and, as before, $\mu \equiv \frac{s^2}{8}\left(\frac{1-\alpha^2}{\alpha^2}\right)$.

To evaluate the mean value of $s$, we return to equation (20):

$$\langle s \rangle = \int_0^\pi \frac{\sin^3\psi \, d\psi}{64\sqrt{2\pi}\alpha} \int_0^\infty s^4 e^{-\frac{s^2}{8}\left(\sin^2\psi + \frac{\cos^2\psi}{\alpha^2}\right)} s \, ds = \int_0^\pi \frac{\sin^3\psi \, d\psi}{64\sqrt{2\pi}\alpha} \times \frac{8^3}{\left(\sin^2\psi + \frac{\cos^2\psi}{\alpha^2}\right)^3} \qquad (22).$$

Again, by writing $x = \cos\psi$, the integral becomes elementary and shields,

$$\langle s \rangle = \sqrt{\frac{2}{\pi}} \left\{ \frac{\alpha(3-2\alpha^2)}{1-\alpha^2} + \frac{3-4\alpha^2}{(1-\alpha^2)^{3/2}} \arccos\alpha \right\} \qquad (23).$$

Thus, from equation (21)

$$F(\mathfrak{R},\alpha) = \langle s \rangle \frac{(\langle s \rangle \mathfrak{R})^4 e^{-\frac{(\langle s \rangle \mathfrak{R})^2}{8}}}{64\sqrt{2\pi}\alpha} \left\{ \frac{e^{-\hat{\mu}}}{\hat{\mu}} + \frac{\sqrt{\pi}(2\hat{\mu}-1) Erf(\sqrt{\hat{\mu}})}{2\hat{\mu}^{3/2}} \right\} \qquad (24),$$

with $\hat{\mu} \equiv \frac{s^2}{8}\left(\frac{1-\alpha^2}{\alpha^2}\right)$ and $\langle s \rangle$ given by (23).

Figure 2 shows graphs of the corresponding pdfs.

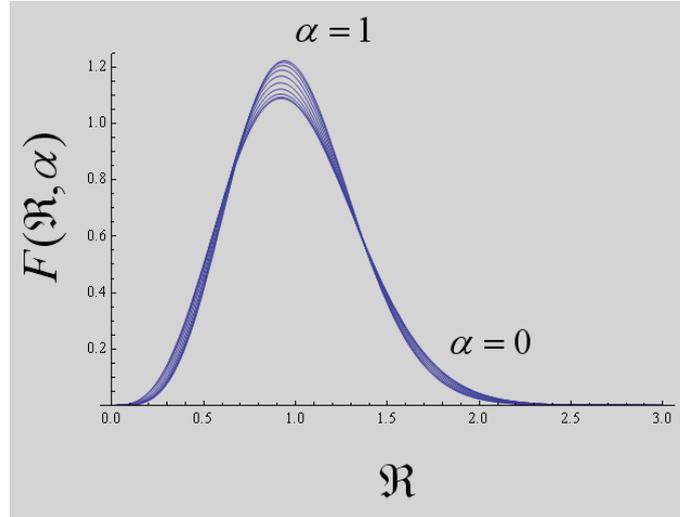

Figure 2. The transitional Ginibre to GSE pdfs $F(\mathfrak{R},\alpha)$ have norm and mean one. Here shown are the pdfs for values of $\alpha$ ranging from 0 to 1.

**Transitional pdf Ginibre To GOE**

We effect such transition $\vec{\alpha} = (000) \rightarrow (110)$ by writing $\vec{\alpha} = \alpha(110)$ and letting $0 \leq \alpha \leq 1$, such that $s = 2\sqrt{g^2 + c^2 + \alpha^2 d^2 + \alpha^2 e^2}$. In this case, the associated space associated with equation (4) is 4-dimensional and the probability density becomes (after performing the integration in $\varphi$):

$$F(s,\alpha) = \frac{s^3}{32\pi\alpha^2} \int_0^\pi \int_0^\pi e^{-\frac{s^2}{8}\left(\sin^2\psi \sin^2\theta + \frac{\sin^2\psi\cos^2\theta + \cos^2\psi}{\alpha^2}\right)} \sin^2\psi \sin\theta\, d\psi d\theta \qquad (25).$$

The expression is integrable in $\theta$ by putting $x = \cos\theta$ then,

$$F(s,\alpha) = \frac{s^3}{32\pi\alpha^2} \frac{2\sqrt{2\pi}\alpha}{s\sqrt{1-\alpha^2}} \int_0^\pi e^{-\frac{s^2}{8\alpha^2}\left(\alpha^2 \sin^2\psi + \cos^2\psi\right)} Erf\left(\frac{s\sqrt{1-\alpha^2}\sin\psi}{2\sqrt{2}\alpha}\right) \sin\psi\, d\psi \qquad (26).$$

We were not able to integrate the expression above exactly. However, good progress is achieved by noticing that $F(s,\alpha)$ is the product of a simple an explicit function of $\alpha$ and $s$ and a parametric integral in the lumped variable $\frac{s\sqrt{1-\alpha^2}}{2\sqrt{2}\alpha}$,

$$F(s,\alpha) = \frac{s^3}{32\pi\alpha^2} \frac{2\sqrt{2\pi}\alpha\, e^{-\frac{s^2}{8}}}{s\sqrt{1-\alpha^2}} \Im(\xi) \qquad (27),$$

where $\xi \equiv \frac{s\sqrt{1-\alpha^2}}{2\sqrt{2}\alpha}$, and

$$\Im(\xi) = \int_0^\pi e^{-\xi^2 \cos^2\psi} Erf(\xi \sin\psi) \sin\psi\, d\psi \qquad (28).$$

From the domains of $s$ and $\alpha$, the new variable can take values $0 \le \xi < +\infty$, then, for convenience we introduce the variable $y$ as $\xi = \frac{y}{1-y}$, such that $0 \le y \le 1$ (which will permit a simpler study of the function).

We found an accurate representation of $\Im(y)$, to within 1% in $0 \le y \le 1$,

$$\Im(y) = y(1-y)\sum_n a_n T_{2n}(y) \qquad (29),$$

where $T_m(y)$ are the Chebyshev polynomials of order m of the first kind, and $a_n$ are constant coefficients, $a_0 = 2.300$, $a_1 = -0.997$, $a_2 = -0.284$, $a_3 = 0.118$, $a_4 = -0.225$, $a_5 = -0.100$.

Surprisingly, the mean $<s>$ can be obtained exactly if we return to equation (25) and first integrate the right-hand side times $s$ from $0$ to $\infty$. The result is,

$$\langle s \rangle = \int_0^\pi \int_0^\pi \frac{3\alpha^3 \sin\theta \sin^2\psi \, d\psi \, d\theta}{\sqrt{2\pi}\left[\cos^2\psi + (\cos^2\theta + \alpha^2 \sin^2\theta)\sin^2\psi\right]^{5/2}} \qquad (30)$$

The integral in $\theta$ can, like before, be obtained by a simple change of variables, then,

$$\langle s \rangle = \int_0^\pi \frac{4\sqrt{\frac{2}{\pi}}\alpha^3 \sin^2\psi \left(2 + \cos^2\psi + \alpha^2 \sin^2\psi\right) d\psi}{\left[1 + \alpha^2 + (1-\alpha^2)\cos 2\psi\right]^2} \qquad (31)$$

which is

$$\langle s \rangle = \sqrt{2\pi}\,\frac{1+\alpha+\alpha^2}{1+\alpha} \qquad (32).$$

Figure 3 shows the pdfs for $0 \leq \alpha \leq 1$. This graph is qualitatively different from those shown in figures 1 and 2. This difference is most noticeable for values of $\Re < 1$, and it is due to the linear to cubic transition of the pdfs in the neighborhood $\Re \approx 0$.

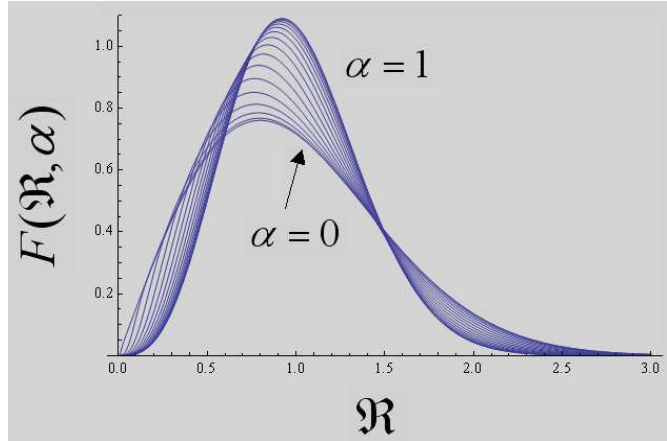

Figure 3. Transitional pdf for GOE to Ginibre.

**Conclusions**

In this article, we have found explicit analytical expressions for some transitional pdfs between some Wigner surmises of RMT and Ginibre. Besides deriving them analytically, we also have checked their values against direct numerical statistical studies of expression (3). The results coincide so well that we did not find reasons to plot comparisons graphs.

These analytical expressions are of practical interest when investigating the statistics of processes that do not exactly fall into one of the Wigner or Ginibre forms. The possibility of finding transitional behavior, might shed light into the underlying physics of the process.

The three transitional pdfs obtained here suggest the existence of a fixed point in all the pdfs, the point of inflection to the right of the maximum. We are looking at this in more detail, first to see if it is true and second, to understand if it has any interesting implications. In addition, we are trying to extend the results obtained in this paper to any value of $\bar{\alpha}$.

**Acknowledgments**. Work supported by Gamson Fund and ANII.

**References**

[1] Wishart J 1928 Generalized product moment distribution in samples Biometrika A **20** 32
[2] Wigner E P 1951 On the statistical distribution of the widths and spacings of nuclear resonance levels Proc. Camb. Phil. Soc. **47** 790
[3] Wigner E P 1955 Characteristic vectors of bordered matrices with infinite dimensions Ann. Math. **62** 548
[4] Wigner E P 1957 Statistical properties of real symmetric matrices with many dimensions Can. Math. Congr. Proc. (Toronto: University of Toronto Press) p 174
[5] Dyson F J 1962 Statistical theory of the energy levels of complex systems, I J. Math. Phys. **3** 140
[6] Dyson F J 1962 Statistical theory of the energy levels of complex systems, II J. Math. Phys. **3** 157
[7] Dyson F J 1962 Statistical theory of the energy levels of complex systems, III J. Math. Phys. **3** 166
[8] Dyson F J 1962 Brownian motion model for the eigenvalues of a random matrix J. Math. Phys. **3** 1191
[9] Dyson F J 1962 The threefold way. Algebraic structure of symmetry groups and ensembles in quantum mechanics J. Math. Phys. **3** 1200
[10] Dietz B, Haake F 1990 Taylor and Padé analysis of the level spacing distributions of random-matrix ensembles Z. Phys. B-Condensed Matter **80** 153
[11] Dietz B, Zyczkowski K 1991 Level-spacing distributions beyond the Wigner surmise Z. Phys. B-Condensed Matter **84** 157
[12] Bleher P, Shiffman B and Zelditch S 2000 Universality and scaling between zeros on complex manifolds Invent. Math. **142** 351
[13] Brody T A, Flores J, French J B, Mello P A, Pandey A and Wong S S M 1981 Random-matrix physics: spectrum and strength fluctuations Rev. Mod. Phys. **53** 385
[14] Kota V K B and Sumedha S 1999 Phys. Rev. E **60** 3405
[15] Sakhr J and Nieminen J M 2006 Phys. Rev. E **73** 036201
[16] Muttalib K A, Chen Y, Ismail M E H and Nicopoulos V N 1993 Phys. Rev. Lett. **71** 471
[17] Fyodorov Y V, Khoruzhenko B A and Sommers H-J 1997 Phys. Rev. Lett. **79** 557
[18] Nieminen J M 2009 Eigenvalue spacing statistics of a four-matrix model of some four-by-four random matrices J. Phys. A **42** 035001
[19] Nieminen J M 2007 Gaussian point process and two-by-two random matrix theory Phys. Rev. E **76** 047202
[20] Metha M L 1991 *Random Matrices* 2nd edn (San Diego, CA: Academic)
[21] Haake F 2001 *Quantum Signatures of Chaos* 2nd edn (Berlin: Springer)
[22] Bohigas O 1991 *Chaos and Quantum Physics* ed Giannoni M-J (Amsterdam: Elsevier) p 87
[23] Weeks J R 1985 The shape of space: how to visualize surfaces and three-dimensional manifolds (Marcel Dekker) Ch 14

[24] Abramowitz M and Stegun I 1972 Handbook of mathematical functions with formulas, graphs, and mathematical tables (U.S. Government Printing Office, 10th printing) p 376

[25] Legendre A M 1825 Traité des fonctions elliptiques et des intégrales Eulériens (Imprimerie de Huzard-Courcier, Rue du Jardinet, $N^o$ 12, Paris) p 256

[26] Haznadar Z and Zeljko Š 2000 Electromagnetic fields, waves and numerical methods (IOS Press) Appendix F, p 403